\documentclass[namedreferences]{solarphysics}
\usepackage[optionalrh]{spr-sola-addons} 
\usepackage{graphicx}        
\usepackage{natbib}         
\usepackage{color}           
\usepackage{url}             

\newcommand{\aap}{    {\it Astron. Astrophys.}}

\newcommand{\aapr}{   {\it Astron. Astrophys. Rev.}}

\newcommand{\apj}{    {\it Astrophys. J.}}
\newcommand{\apjl}{   {\it Astrophys. J. Lett.}}

\newcommand{\solphys}{{\it Solar Phys.}}

\newcommand{\ssr}{    {\it Space Sci. Rev.}}

\begin{document}

\begin{article}

\begin{opening}

\title{LOFAR observations of fine spectral structure dynamics in type IIIb radio bursts}

\author{I.\,N.~\surname{Sharykin}$^{1,2,3}$\sep
        E.\,P.~\surname{Kontar}$^3$\sep
        A.\,A.~\surname{Kuznetsov}$^{2}$\sep
       }
\runningauthor{Sharykin et al.}
\runningtitle{LOFAR observations of type IIIb bursts}

   \institute{$^{1}$ Space Research Institute (IKI), Moscow, Russia
                     email: \url{ivan.sharykin@phystech.edu}\\
					$^{2}$ Institute of Solar-Terrestrial Physics, Irkutsk, Russia \\
                    $^{3}$ Glasgow University, School of Astronomy and Astrophysics, Glasgow, UK
             }

\begin{abstract}
Solar radio emission features a large number of fine structures demonstrating great variability in frequency and time. We present spatially resolved spectral radio observations of type IIIb bursts in the $30-80$ MHz range made by the Low Frequency Array (LOFAR). The bursts show well-defined fine frequency structuring called ``stria'' bursts. The spatial characteristics of the stria sources are determined by the propagation effects of radio waves; their movement and expansion speeds are in the range of $(0.1-0.6)c$. Analysis of the dynamic spectra reveals that both the spectral bandwidth and the frequency drift rate of the striae increase with an increase of their central frequency; the striae bandwidths are in the range of $\sim (20-100)$ kHz and the striae drift rates vary from zero to $\sim 0.3$ MHz $\textrm{s}^{-1}$. The observed spectral characteristics of the stria bursts are consistent with the model involving modulation of the type III burst emission mechanism by small-amplitude fluctuations of the plasma density along the electron beam path. We estimate that the relative amplitude of the density fluctuations is of $\Delta n/n\sim 10^{-3}$, their characteristic length scale is less than 1000 km, and the characteristic propagation speed is in the range of $400-800$ km $\textrm{s}^{-1}$. These parameters indicate that the observed fine spectral structures could be produced by propagating magnetohydrodynamic waves.
\end{abstract}
\end{opening}

\section{Introduction}
During solar flares beams of accelerated electrons propagating
along the open magnetic field lines can produce the so-called type III radio bursts \citep{2008LRSP....5....1B,Holman2011}. They are observed as fast drifting structures in radio dynamic spectra with high brightness temperature \citep[see][as recent reviews]{1985srph.book..289S,Pick2008,2008LRSP....5....1B}. The type III radio bursts can be traced from the solar corona into the interplanetary space \citep[e.g.][]{1974SSRv...16..189L,2011SoPh..273..413K,Krupar2014,2015A&A...582A..52A}, where the corresponding electron beams and the beam-driven Langmuir waves
near the electron plasma frequency $f_{\mathrm{pe}}$ can be observed in-situ \citep{Lin1985, Krucker2007}. In the dynamic spectra of type III bursts,the fundamental ($\approx f_{\mathrm{pe}}$) and harmonic ($\approx 2f_{\mathrm{pe}}$) parallel drifting components are usually identified \citep{McLean1985}.

A large fraction of metric and decametric type III radio bursts reveal fine spectral structuring. In particular, the so-called type IIIb radio bursts \citep{deLaNoe1972,1979SoPh...62..145A,2017SoPh..292..155M} are characterized by multiple narrowband bursts with slow frequency drift, known as stria bursts. They  compose the fast drifting spectral structure similar to usual type III radio bursts \citep{Ellis1967, Ellis1969, deLaNoe1972, Stewart1975, Takakura1975, Baselyan1974a}. Striae can be observed in both the fundamental and harmonic emission components, although the harmonic striae are more diffusive \citep{Baselyan1974b}; if present, they can form the so called IIIb--IIIb pairs \citep{1979SoPh...62..145A,2015RRPRA..20...99B}. A typical stria frequency bandwidth is of about $\sim 30-300$ kHz with the frequency drift rate of $0-150$ kHz $\textrm{s}^{-1}$ \citep{Bhonsle1979, Kruger1984}. Duration of an individual stria depends on its frequency and is of about $\sim 1$ second for the fundamental component and can be several time longer for the harmonic emission \citep{Bhonsle1979, Kruger1984}.

The basic explanation of striae origin is the existence of density variations along the electron beam path; this idea was firstly proposed by \inlinecite{Takakura1975}. Numerical modelling of \inlinecite{Kontar2001} demonstrated that the spatial distribution of Langmuir waves is strongly modulated by small-amplitude density fluctuations creating ``Langmuir wave clumps'' that could be responsible for individual striae. Estimations in that work showed that even relatively weak density perturbations ($\Delta n/n\sim 10^{-3}$, where $n$ is the thermal electron density) are sufficient to form the observed fine spectral structures.  However, it is still not clear which magnetohydrodynamic (MHD) waves cause these density fluctuations; there is a long list of possible mechanisms responsible for the emission modulation \citep{Melrose1982}. Furthemore, until now the number of studies devoted to investigation of striae spectral properties at different frequencies was limited, while a successful theory should be able to explain both the observed striae drift rates and bandwidths as well as their frequency dependencies.

In this paper, we analyze two IIIb bursts observed with the LOw Frequency ARray \citep[LOFAR,][]{Haarlem2013} on 16 April 2015. The frequency resolution of LOFAR is sufficient to resolve striae; it allows also studying the spatial characteristics of the emission sources. While the spatial dynamics of the striae sources in the mentioned event was analyzed in detail in the paper of \inlinecite{Kontar2017}, the main aim of this paper is to investigate the spectral characteristics of the striae (i.e., bandwidth and frequency drift) and their dependencies on the emission frequency. Our particular interest is to verify applicability of the density fluctuations model to explaining the observed striae properties.

\section{LOFAR observations}
LOw Frequency ARray \citep[LOFAR,][]{Haarlem2013} was designed by the Netherlands Institute for Radio Astronomy (ASTRON). It is working in the metric-decametric wavelengths range and is able to produce spatially resolved solar observations with high time cadence and excellent spectral resolution. In this work we analyse the low-band observations (in the 30-80 MHz range) made with the LOFAR core stations located near Exloo, Netherlands; all 24 core stations (scattered over the area of $\sim 3\times 2$ $\textrm{km}^2$) were used for the observations.

Instead of a classical interferometric approach, the spatially-resolved spectroscopic observations were performed using the LOFAR beam-formed mode \citep{Stappers2011, Haarlem2013}, when the data from the LOFAR core stations are combined to form a number of ``tied-array beams'' covering an area of the sky; the beam size is of about $\lambda/D\sim 10'$ at 32 MHz, where $\lambda$ is the wavelength and $D$ is the maximum baseline. The advantage of using the tied-array beams is that it allows producing images with very high time resolution, which is not possible in the LOFAR interferometric mode \citep[see][for details]{Stappers2011,Morosan2014,Morosan2015,2017A&A...606A.141R,2018ApJ...856...73C}.

In this work the LOFAR configuration included 127 beams covering the solar disk and adjacent areas with the separation between the beam centers of about $356''$ (see Figure \ref{ims}). The fluxes corresponding to each beam were recorded with high frequency and time resolution ($12.2$~kHz and 10 ms, respectively). Flux calibration
was made using observations of the Crab nebula (Tau A) as it was demonstrated by \citet{Kontar2017,2018ApJ...856...73C}. In the below light curves and dynamic spectra, the pre-event (pre-burst) background is subtracted.

\section{Main properties of the selected events}
\subsection{LOFAR dynamic spectra of the selected type IIIb radio bursts}\label{DSp}
\begin{figure}
\centering
\includegraphics[width=1.0\linewidth]{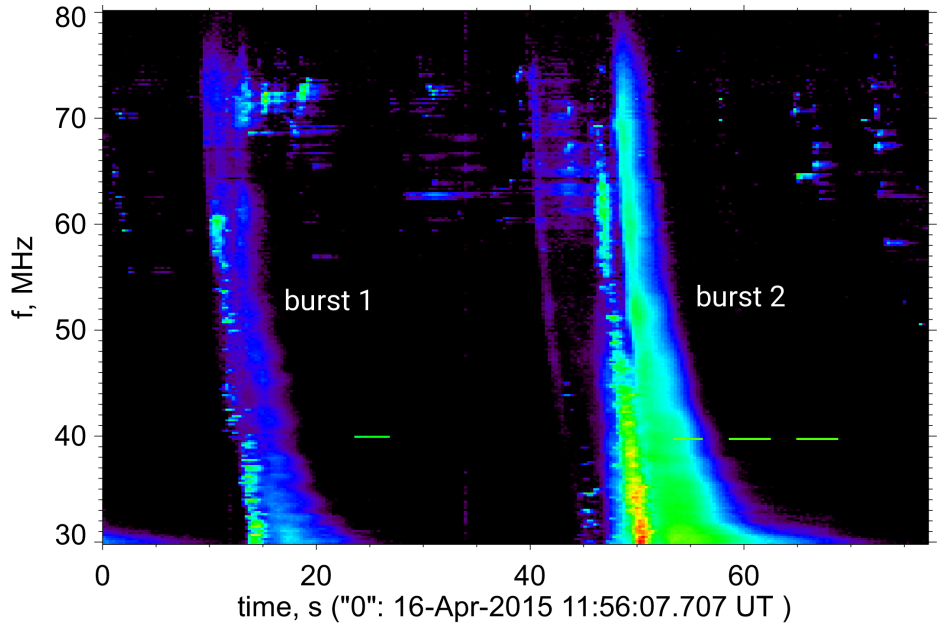}
\caption{Spatially-integrated background-subtracted calibrated LOFAR dynamic spectrum of two subsequent type IIIb radio bursts.}
\label{dyn_spec}
\end{figure}

The observations were made on 16 April 2015, around the local noon; a number of type III bursts were detected. Among them, we selected two bursts shown in Figure \ref{dyn_spec}; these bursts were chosen because they are isolated, i.e., do not overlap with other bursts. The bursts occurred at $\sim$11:56:20 and $\sim$11:56:55 UT; below, they are referred to as burst 1 and burst 2, respectively. We note that there were no flares or coronal mass ejections during the considered time interval; the nearest flare (of C2 GOES class) was at $\sim$10:45:00~UT. Both bursts reveal the typical two-component (fundamental-harmonic) structure of dynamic spectra; the fine spectral structures (striae) are well visible in both of them. Below we demonstrate in detail the analysis technique and results for burst 2 (the brighter one); the other one is analyzed in the same way. The statistical results are shown for both bursts.

\begin{figure}
\centering
\includegraphics[width=1.0\linewidth]{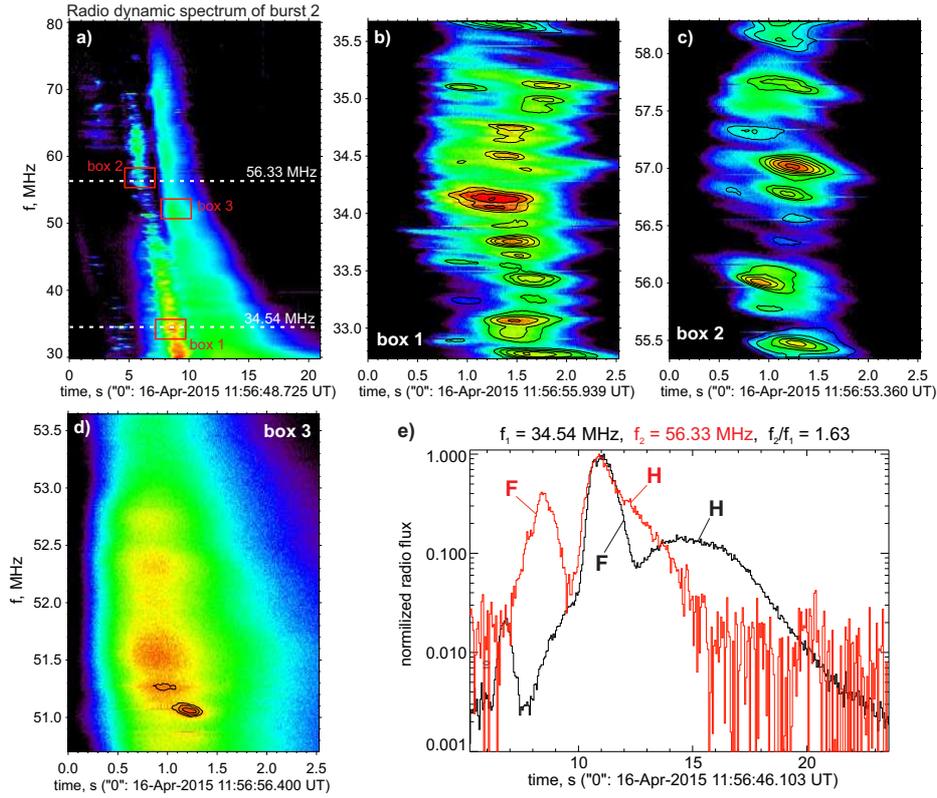}
\caption{a) Dynamic spectrum of burst 2; three rectangular boxes show the regions of dynamic spectrum presented in other panels. b-d) Zoomed regions of the dynamic spectrum corresponding to the fundamental (b-c) and harmonic (d) components; the contours mark striae. e) Light curves at two selected frequencies demonstrating the fundamental-harmonic relation.}
\label{zoom}
\end{figure}

In Figure \ref{zoom} we show zoomed dynamic spectra of burst 2 for the chosen frequency and time ranges. These ranges are marked by rectangular boxes in top left panel of the figure. As said above, the burst is composed of two distinct components: a bright (fundamental) component with well-pronounced striae is followed by a more diffusive drifting structure (harmonic); hence the analyzed bursts can be identified as type IIIb--III pairs. The drift rates of both components are comparable and are of about 10 MHz $\textrm{s}^{-1}$ that corresponds to the electron beam speed of $\sim 0.3c$ assuming the one-fold Newkirk coronal density model \citep{Newkirk1961}. The dynamic spectra of the diffusive component (harmonic) reveal fine spectral structures, too (see box 3 in Figure \ref{zoom}); these structures look like smoothed striae. Below, we analyze in detail the striae detected in the first bright (fundamental) component since they are more distinctive (see boxes 1 and 2 in Figure \ref{zoom}).

As an example, we show in the right-bottom panel of Figure \ref{zoom} the time profiles of the total (spatially-integrated) radio flux at two frequencies: $f_1=34.54$ MHz and $f_2=56.33$ MHz; the frequencies were chosen to ensure that the corresponding radio fluxes (i.e., the fundamental and harmonic components) peak at the same time. Therefore the harmonic-fundamental ratio at this particular time can be estimated as $f_2/f_1\approx 1.63$, which is relatively low but not unprecedented for type III bursts \citep{1985srph.book..289S}.

In panels (b-d) of Figure \ref{zoom}, individual striae with high contrast are highlighted by contours; the contour levels are chosen to ensure that the contours are not too long (with the length below a threshold of 250 pixels of dynamic spectrum) and hence they, as a rule, enclose separate striae rather than groups of striae.
The stria durations vary in the range of $0.5-1$ s. One can notice that the striae have frequency dependent drift rate: the stria drift rate is larger at higher frequencies; this behaviour is typical of the type III radio bursts in general. Also, the high-frequency striae have a slightly larger bandwidth than the low-frequency ones. In Section \ref{stat_spec}, these frequency variations of the stria parameters are analyzed quantitatively.

\subsection{Dynamics of the radio emission source}
\begin{figure}
\centering
\includegraphics[width=1.0\linewidth]{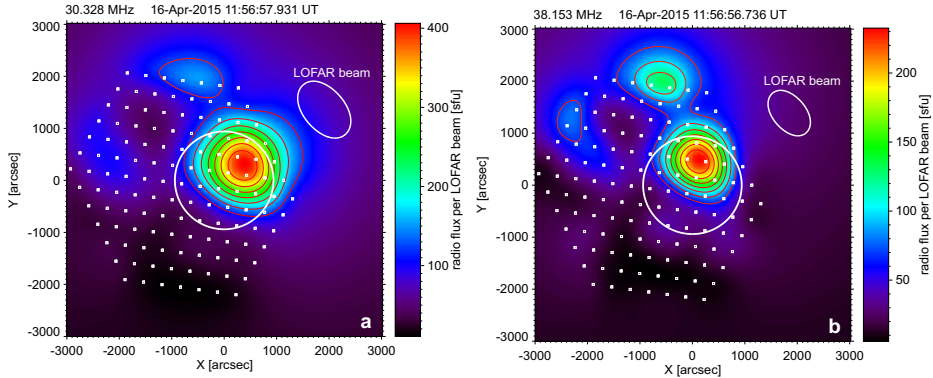}
\caption{LOFAR radio images at two time-frequency points within burst 2. Centers of 127 LOFAR beams are shown by small white squares. The solar limb is shown by white thick circle. The LOFAR beams sizes (at $1/2$ level) at the considered frequencies are shown by white ellipses.}
\label{ims}
\end{figure}

Figure \ref{ims} shows two examples of the LOFAR radio images (at different times and frequencies) obtained by interpolation of the fluxes corresponding to different beams using the radial basis function method. At low frequencies, we can see a single well-defined radio emission source located on the solar disk; the signal-to-noise ratio is remarkably high. At higher frequencies (see, e.g., right panel in Figure \ref{ims}), additional weaker and smaller sources appear that are likely caused by the instrument sidelobes. For these reasons, we analyze the spatial characteristics of the emission source up to the frequency of $\approx 50$ MHz only; on the other hand, the spectral characteristics of the striae are analyzed up to 70 MHz. From Figure \ref{ims}, one can notice that with an increase of the emission frequency the emission source shifts eastward, and its size decreases. The raw images do not allow us to inspect dynamics of the source position with a sufficient accuracy; therefore, to describe the source parameters quantitatively and to study temporal dynamics, we fitted it by an elliptical Gaussian defined by seven free parameters: normalization coefficient, center position ($x_0$ and $y_0$), size ($\sigma_x$ and $\sigma_y$), and tilt angle. Both the parameters and their confidence limits (errors) were estimated using the least-squares procedure. Only the LOFAR beams located within $1000''$ from the solar disk center were used in the fitting procedure.

We should note that the observed radio map is a convolution of the real brightness distribution with the LOFAR beam. In a simple case when both the emission source and the LOFAR beam have approximately Gaussian shapes, the real source area $A_{\mathrm{real}}$ is determined as $A_{\mathrm{real}} \approx A_{\mathrm{obs}} - A_{\mathrm{beam}}$ where $A_{\mathrm{obs}}$ and $A_{\mathrm{beam}}$ are the areas of the observed emission source and the LOFAR beam, respectively. Thus, firstly, the apparent increase of emission source size with decreasing frequency seems to be caused mainly by the LOFAR beam broadening with decreasing frequency. \inlinecite{Kontar2017} estimated the real source size for the considered event as $17-22$ arcmin around 32~MHz, which is approximately twice larger than the LOFAR beam at that frequency. Secondly, the expansion/shrinking rate of the emission source at a fixed frequency (which is applicable to a narrowband stria) satisfies the relation $\mathrm{d}A_{\mathrm{real}}/\mathrm{d}t = \mathrm{d}A_{\mathrm{obs}}/\mathrm{d}t$ (because the LOFAR beam size at a fixed frequency is constant). That is why the expansion rates of individual striae (considered below) are not affected by the mentioned convolution effect.

In addition, in the observed frequency range the ionospheric refraction can result in displacement of the apparent radio emission source. LOFAR monitoring of the point-like source Tau~A revealed absence of significant intensity scintillations on the subsecond time scales; the level of ionospheric turbulence at the time of observations was low. To minimize the (possible) ionospheric effects, we focus on relative motions in the plane of sky at single frequency at the timescales shorter than the ionospheric scintilations observed.

\begin{figure}
\centering
\includegraphics[width=1.0\linewidth]{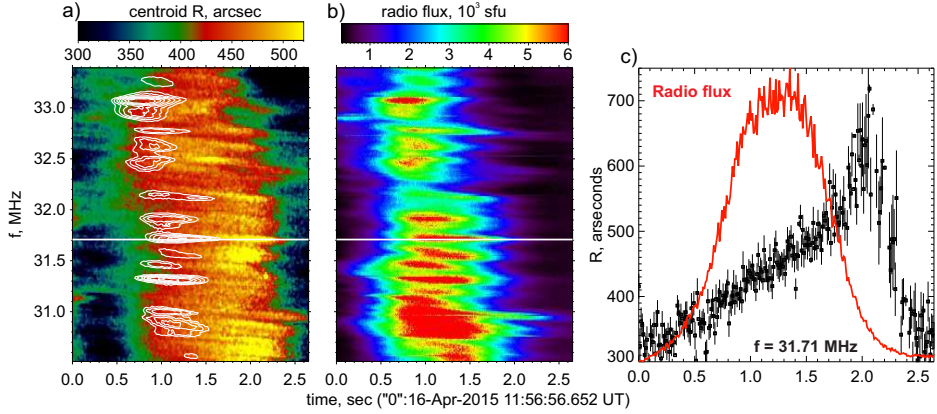}
\caption{Dynamics of the emission source position on the solar disk (for a sub-region of burst 2, fundamental component). a) Radial distance from the solar disk center shown by colored background as function of time and frequency; the emission intensity contours (white) are overplotted to mark striae. b) The corresponding dynamic spectrum. c) Time profile of the mentioned radial distance at a fixed frequency (31.71 MHz); the intensity lightcurve (in relative units) is overplotted.}
\label{RST}
\end{figure}

Figure \ref{RST} presents comparison of fine structures seen in dynamic spectra with the radial shift of the emission source position, for a region of the dynamic spectrum corresponding to the fundamental component of burst 2; the time resolution is 12.5 ms. The radial position of the source is calculated as offset of Gaussian centroid from the solar disk center; in panel~a of Figure \ref{RST}, the source position is shown in a dynamic-spectrum style (by colored background). Panel~b presents corresponding region of the dynamic spectrum. In the panel~c the dynamics of the radial distance at a chosen single frequency (31.71 MHz) is shown. The emission intensity is overplotted by white contours (in panel~a) or red lightcurve (in panel~c). Both the time-frequency plots (for all striae) and single-frequency time profiles demonstrate a complicated pattern reported earlier by \inlinecite{Kontar2017}: the emission source position is characterized by a gradually increasing radial distance (i.e., motion towards the limb) with a subsequent fast return motion. The evolution of the source position is delayed with respect to the intensity enhancement; the maximum radial distance is achieved $\sim 1$ s after the intensity peak, consistent with the conclusion by \inlinecite{Kontar2017} that this behaviour most likely reflects the radio emission propagation effects.

\subsection{Parameters of individual striae}\label{case_study}
\begin{figure}
\centering
\includegraphics[width=1.0\linewidth]{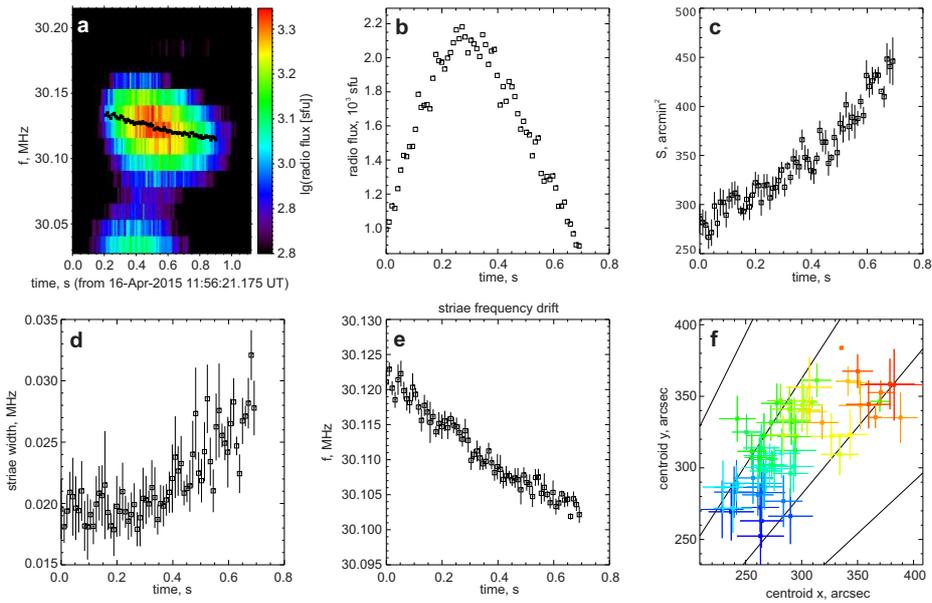}
\caption{An example of ``low-frequency'' stria. a) Dynamic spectrum. Black dots mark the central frequency of the stria (obtained by Gaussian fitting) in each time bin. b) Emission intensity (at the central frequencies) vs. time. c) Area of the radio emission source (at the central frequencies) vs. time. d) Spectral bandwidth of the stria vs. time. e) Central frequency of the stria vs. time. f) Source position on the solar disk at different times (color-coded, with the time increasing from violet to red); black lines show radial directions from the disk center.}
\label{stria1_b1}
\end{figure}

\begin{figure}
\centering
\includegraphics[width=1.0\linewidth]{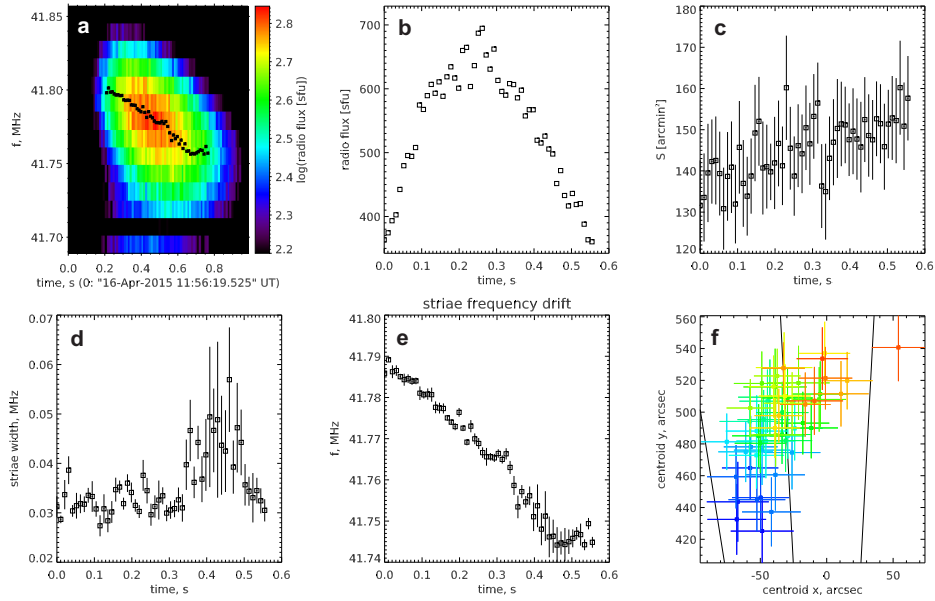}
\caption{Same as in Figure \protect\ref{stria1_b1}, for a ``high-frequency'' stria.}
\label{stria2_b1}
\end{figure}

Figures \ref{stria1_b1}--\ref{stria2_b1} present examples of individual striae. We have selected two striae within burst 1, with the frequencies around 30.11 MHz (Figure \ref{stria1_b1}) and 41.77 MHz (Figure \ref{stria2_b1}); below in this Section, we refer to them as to the ``low-frequency'' (LF) and ``high-frequency'' (HF) striae, respectively. To determine the stria parameters, we fitted the emission spectrum (containing several spectral channels which were selected manually for each stria burst) in each time bin by a Gaussian with a linear background. This provides us with the time-dependent values of the stria central frequency and bandwidth (determined at one-sigma level); the radio flux and emission source size and position shown in Figures \ref{stria1_b1}--\ref{stria2_b1} correspond to the mentioned (variable) central frequency.

One can see that the emission source experiences gradual expansion with time: from 270 to 450 $\textrm{arcmin}^2$ for the LF stria and from 135 to 155 $\textrm{arcmin}^2$ for the HF stria. Also, both considered striae demonstrate motion of the emission sources in a mostly radial direction, similar to the behaviour observed at a fixed frequency (see Figure \ref{RST}).

The real emission source size can be estimated using the above-mentioned relation $A_{\mathrm{real}} \approx A_{\mathrm{obs}} - A_{\mathrm{beam}}$, from which we estimate the linear source size (FWHM) in the plane of sky as $l_{\mathrm{real}}\approx 2\sigma_{\mathrm{real}}\sqrt{2\ln 2}\approx (2\sqrt{2\ln 2}/\pi) \sqrt{A_{\mathrm{obs}}-A_{\mathrm{real}}}\approx 0.75\sqrt{A_{\mathrm{obs}}-A_{\mathrm{real}}}$. For the LF stria peak, considering $A_{\mathrm{real}} \approx 320$~arcmin$^2$ and $A_{\mathrm{beam}} \approx 110$~arcmin$^2$, we obtain $l_{\mathrm{real}} \approx 11$ arcmin; for the HF stria peak, we have $A_{\mathrm{real}} \approx 145$~arcmin$^2$, $A_{\mathrm{beam}} \approx 60$~arcmin$^2$, and $l_{\mathrm{real}} \approx 7$ arcmin.

Spatial extent of the source along the line-of-sight (including the effects of scattering) can be estimated from FWHM of temporal stria intensity profile in the same way as in the work of \inlinecite{Kontar2017}. For the both LF and HF striae, the duration at a fixed frequency $\Delta t$ is about 0.6 seconds; thus the upper limit of the spatial extent is $(l_{\mathrm{LOS}})_{\max} = c\Delta t\approx 180~\textrm{Mm}\approx 4$~arcmin. The size of the stria emission source along the line-of-sight $l_{\mathrm{LOS}}$ is smaller than the size measured across the line-of-sight; thus, studying separate striae we confirmed the conclusion of \inlinecite{Kontar2017} that the radio wave scattering in the corona is highly anisotropic.

The LF stria bandwidth increases with time from 38 to 54 kHz, while the HF stria reveals no trend in bandwidth evolution. To estimate a characteristic (mean) stria bandwidth, we calculated a cumulative stria spectrum by summation of the spectra in all relevant time bins (shifted to provide the same central frequency). Then the bandwidth of the resulting spectrum (at one-sigma level) is considered to be a characteristic stria bandwidth; it is of about 44 kHz for the LF stria and 68 kHz for the HF stria (the HF stria is wider). Also, the HF stria reveals faster frequency drift compared with the LF stria: the drift rate can be estimated as $-27$ kHz $\textrm{s}^{-1}$ for the LF stria and $-98$ kHz $\textrm{s}^{-1}$ for the HF stria.

A similar analysis was performed for all identified striae in the bursts 1 and 2, with the results presented and summarized in the next Section.

\section{Statistics of the striae parameters}
\subsection{Dynamics of the emission sources}\label{stat_sourc}
\begin{figure}
\centering
\includegraphics[width=1.0\linewidth]{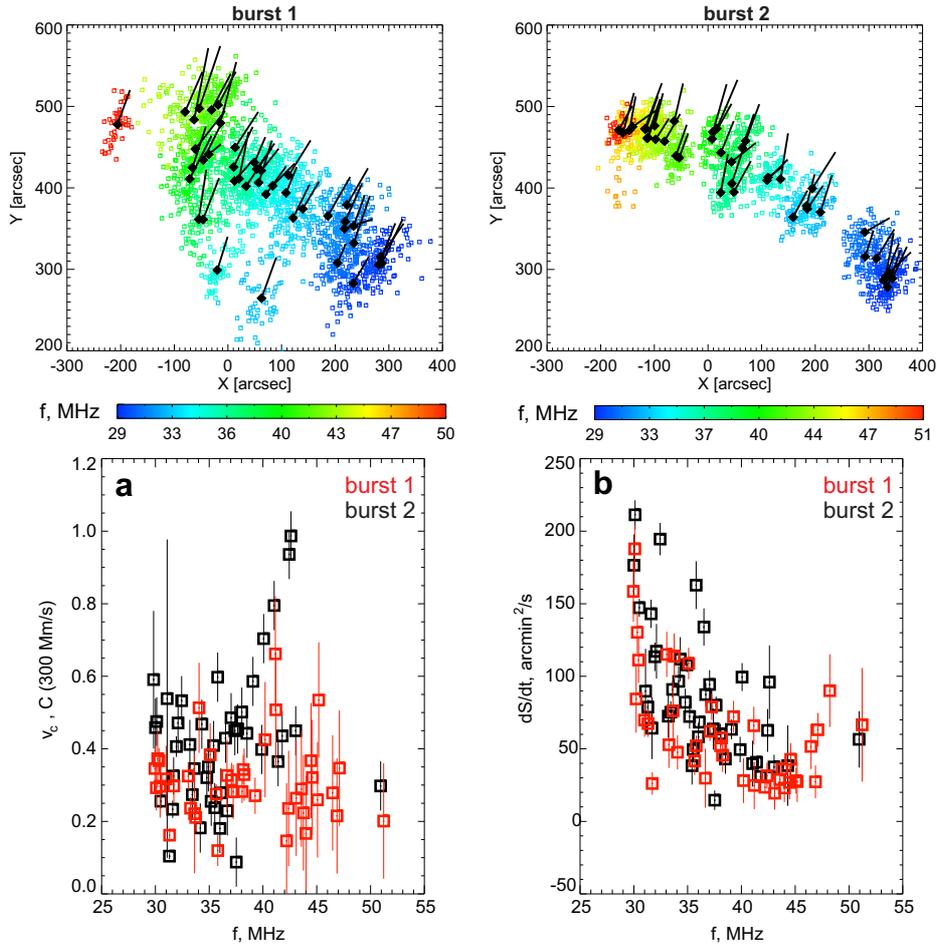}
\caption{Upper panels: central positions of the radio emission sources; color marks the emission frequency and black dots show the average positions of the radio emission sources for different striae. Black lines show the average velocities (direction and relative magnitude) of the striae radio emission sources. Bottom panels: speeds (a) and area expansion rates (b) of the radio emission sources for different striae vs. the striae central frequencies; red and black colors mark bursts 1 and 2, respectively.}
\label{stat_source}
\end{figure}

We have selected for quantitative analysis 43 striae in burst 1 and 40 striae in burst 2. Top panels in Figure \ref{stat_source} show central positions of the radio emission sources for different times and frequencies; the average positions and the characteristic (average) velocities of the emission sources for the selected striae are shown as well. To determine an average vector of the emission source velocity in the plane of the sky, we used linear fits of $x_0(t)$ and $y_0(t)$ dependencies, where $(x_0, y_0)$ are the central coordinates of the emission source; the corresponding average speeds (absolute magnitudes) are shown in the bottom panel (a) of Figure \ref{stat_source}. A similar procedure (linear fitting) was used to determine a characteristic expansion rate of the emission source $\mathrm{d}S/\mathrm{d}t$, see bottom panel (b) of Figure \ref{stat_source}. We do not show the error bars for $x_0$ and $y_0$ in top panels of Figure \ref{stat_source} because this would make the figure unreadable due to the large number of data points; the typical error bars at two representative frequencies are shown in Figures \ref{stria1_b1}f and \ref{stria2_b1}f.

One can see that the obtained speeds of the striae radio emission sources $v_{\mathrm{c}}$ are mostly in the range of $(0.1-0.6)c$ and only a few points are outside this range. The area expansion rate of the emission source can be as high as $\sim 200$ $\textrm{arcmin}^2$ $\textrm{s}^{-1}$. Assuming that the emission source has a roughly circular shape, we can estimate the corresponding linear expansion rate as $\mathrm{d}r_{\bot}/\mathrm{d}t\approx (\mathrm{d}S/\mathrm{d}t)/(2\pi\sqrt{S})$; this value is also comparable with the speed of light and can be as high as $0.2c$. The expansion rate of the emission source tends to decrease with an increase of the emission frequency.

\subsection{Statistics of the striae bandwidths and frequency drift rates}\label{stat_spec}
\begin{figure}
\centering
\includegraphics{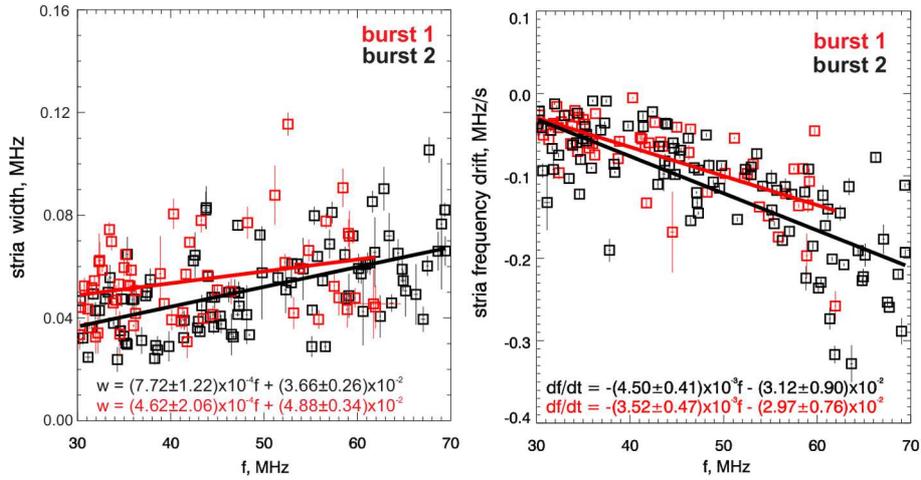}
\caption{Statistics of the striae bandwidths (left) and frequency drift rates (right). Solid lines represent linear fits to the data points. Red and black colors show the results for burst 1 and 2, respectively.}
\label{stat_wdfdt}
\end{figure}

The left panel of Figure \ref{stat_wdfdt} shows the striae bandwidths calculated according to the technique described in Section \ref{case_study}. The resulting data points are fitted by linear functions (separately for the bursts 1 and 2) with the equations written within the plot. The striae bandwidth increases with an increase of frequency for both bursts; typical striae widths at 30 and 60 MHz are of about 40 and 60 kHz, respectively (i.e., just a few LOFAR frequency channels). The relative bandwidth $\Delta f/f$  is weakly variable being of about 0.13\% and 0.10\% at the mentioned frequencies. We can conclude that the linear fits for both type IIIb bursts are similar to each other, and an average stria bandwidth increases with frequency by $\sim 0.6$ kHz per MHz.

The frequency drift rate of the striae increases with an increase of the emission frequency, too (see the right panel of Figure \ref{stat_wdfdt}); it varies from $\sim 30$ kHz $\textrm{s}^{-1}$ at 30 MHz up to $\sim 150$ kHz $\textrm{s}^{-1}$ at 60 MHz. The obtained linear fits (stria drift rate vs. frequency) for both considered type IIIb bursts are similar to each other: $\mathrm{d}f/\mathrm{d}t\simeq 0.004f$. Note that the striae drift rates are much smaller than the typical drift rate of usual type III bursts, which normally
decreases with frequency as $\mathrm{d}f/\mathrm{d}t\simeq 0.01 f^{1.84}$ \citep{1973SoPh...29..197A}.

\section{Discussion}
The most straightforward scenario for the striae formation in the type III bursts is an existence of plasma density fluctuations along the electron beam path \citep{Takakura1975}. These small-amplitude density perturbations can modulate the Langmuir waves generation substantially \citep{Kontar2001} and hence produce fine structures like striae in the dynamic radio spectra.

\begin{figure}
\centering
\includegraphics[width=0.8\linewidth]{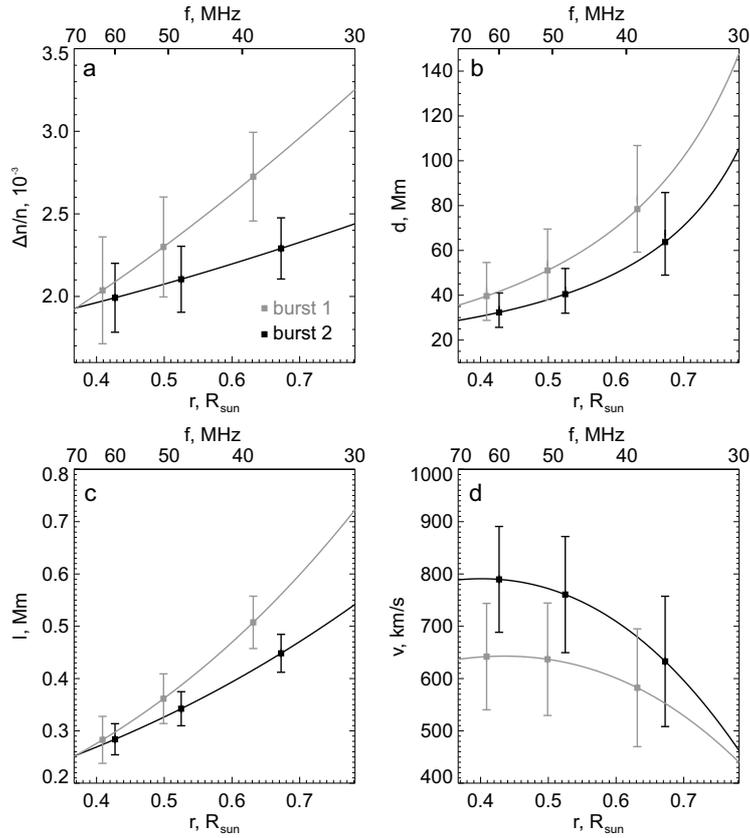}
\caption{Various parameters inferred from the striae bandwidths, durations, and drift rates: a) amplitude of the plasma density perturbations $\Delta n/n$; b) characteristic electron beam length $d$; c) characteristic length scale of the plasma inhomogeneities $l$; d) propagation speed of the plasma density perturbations. The respective heights are calculated according to the Newkirk coronal density model.}
\label{esti}
\end{figure}

Results obtained in this work (see Section \ref{stat_spec}) can be used to  estimate the properties of the density irregularities that are responsible for striation. Assuming that the emission is produced at the local plasma frequency, we estimate the relative density variations corresponding to the stria bursts as $\Delta n/n\simeq 2\Delta f/f$, where $\Delta f$ is bandwidth of a stria and $f$ is its central frequency. From the linear fits in the right panel of Figure \ref{stat_wdfdt}, we determined $\Delta n/n$ at different frequencies (see Figure \ref{esti}a) for both type IIIb bursts. The amplitude of the density fluctuations varies from $\sim 2.0\times10^{-3}$ at 70 MHz to $\sim 3.2\times 10^{-3}$ at 30 MHz; the error bars show possible ranges of $\Delta n/n$ considering errors of linear fitting presented in Figure \ref{stat_wdfdt}.

Longitudinal (i.e., along the magnetic field) size of the electron cloud generating the radio emission can be estimated as $d \sim v_{\mathrm{b}}\tau$, where $v_{\mathrm{b}}$ is the electron beam speed and $\tau$ is the characteristic time of interaction between the electron cloud and a particular density inhomogeneity. We assume that the electron beam speed is $v_{\mathrm{b}}=0.3c$ which is consistent with the estimations obtained in Section \ref{DSp}; the time $\tau$ can be estimated as $\tau \sim \Delta f (\mathrm{d}f/\mathrm{d}t)^{-1}$, where $\Delta f$ is a stria bandwidth and $\mathrm{d}f/\mathrm{d}t$ is its frequency drift rate. Note that we do not consider the stria duration as $\tau$, because the stria duration can be affected by propagation effects \citep{Kontar2017} resulting in a radio echo and extending the radio pulses; therefore the above expression characterizes the intrinsic properties of the radio emission source more accurately. The obtained values of the electron beam size vary from $\sim 20$ Mm to $\sim 150$ Mm and increase with a decrease of the stria frequency (see Figure \ref{esti}b). We interpret this as an expansion of the electron beam during its propagation in the solar corona, which is likely determined by geometry of open magnetic flux tube where the electrons propagate.

The characteristic length scale $l$ of the plasma inhomogeneities producing the striae bursts is related to the striae bandwidth as
\begin{equation}
l\approx 2n\left(\frac{\mathrm{d}n}{\mathrm{d}r}\right)^{-1}\frac{\Delta f}{f}\approx 653\left[\log_{10}\left(\frac{f}{f_0}\right)\right]^{-2}\frac{\Delta f}{f},
\end{equation}
where $n$ is the plasma density and $r$ is the distance along the electron beam path. The second expression in the above formula was obtained by assuming radial propagation of the energetic electrons and the Newkirk coronal density model \citep{Newkirk1961}, with $f_0=1.84$ MHz and the resulting value of $l$ in Mm units. The estimated values of $l$ vary in the range of $\sim (0.2-0.8)$ Mm; they increase with a decrease of the stria frequency (see Figure \ref{esti}c). These inhomogeneities are much smaller than the electron beam: $l/d\sim (0.2-2.6)$\%.

Similarly, the characteristic propagation speed $v_{\mathrm{p}}$ of the plasma inhomogeneities is related to the striae frequency drift rate as
\begin{equation}
v_{\mathrm{p}}\approx 2n\left(\frac{\mathrm{d}n}{\mathrm{d}r}\right)^{-1}\frac{1}{f}\frac{\mathrm{d}f}{\mathrm{d}t}\approx 6.53\times 10^5\left[\log_{10}\left(\frac{f}{f_0}\right)\right]^{-2}\frac{1}{f}\frac{\mathrm{d}f}{\mathrm{d}t},
\end{equation}
with the resulting value of $v_{\mathrm{p}}$ in the second expression in km $\textrm{s}^{-1}$ units. The estimated values of the propagation speed vary in the range of $400-800$~$\textrm{km~s}^{-1}$ (see Figure \ref{esti}d) that corresponds to typical speeds of MHD waves \citep{2000SoPh..193..139R}; thus, the striae frequency drift can be caused by the motion of the plasma density perturbations due to MHD
waves \citep[e.g.][for details]{2018arXiv180508282K}.
These perturbations appear supersonic with the speeds which are $2-4$ times larger than a typical sound speed of $c_{\mathrm{s}}\approx 147\sqrt{T/[1~\mathrm{MK}]}\approx 200$ km $\textrm{s}^{-1}$ for a typical coronal plasma temperature of $T\approx 2$ MK \citep{2005psci.book.....A}.

\section{Conclusion}
We presented detailed analysis of spatially-resolved multi-frequency LOFAR observations of two type IIIb radio bursts. The results obtained provide statistically significant properties of individual striae and hence essential constraints for the theories describing the fine spectral structure of type IIIb bursts.
The main results can be summarized as follows:
\begin{itemize}
\item
Spatial position of the radio emission sources is characterized by radial motion from the Sun centre in the sky plane. The motion is particularly well pronounced during the decay phase of a striae.
\item
Evolution of the spatial source position is delayed by $\sim 1$ s with respect to the radio intensity time profiles.
\item
The apparent speed of the radio emission sources in the plane of sky is $(0.1-0.6)c$; the expansion speed of the sources is up to $\sim 0.2c$.
\item
Instantaneous bandwidth of the striae increases with an increase of the central frequency; the bandwidths lie in the range of $20-100$ kHz that corresponds to the relative bandwidth $\Delta f/f$ of about $0.06-0.12$\%.
\item
Frequency drift rate of the striae increases with an increase of the central frequency; the drift rates lie in the range
of $0-0.3$ MHz~$\textrm{s}^{-1}$.
\item
The relative amplitudes of the plasma density fluctuations
that may be responsible for the formation of the striae should be of about $(2-3)\times 10^{-3}$.
\item
The characteristic sizes and propagation speeds of the mentioned density fluctuations are expected to be of about $200-800$ km and $400-800$ km $\textrm{s}^{-1}$, respectively. The propagation speeds are substantially larger than the typical sound speed of 200 km $\textrm{s}^{-1}$ in the corona and closer to the typical Alfv\'en speed \citep{2008A&A...491..297R}.
\item
Estimations of the apparent radio source size (with account for the scattering effects) indicate that the source size across the line-of-sight exceeds considerably the size along the line-of-sight; this implies that scattering of the radio waves must be anisotropic.
\end{itemize}

The results obtained from analysis of two type IIIb bursts with striae support the conclusion of \inlinecite{Kontar2017} that the sizes of the radio emission sources and their dynamics are determined by the radio wave propagation effects. At the same time, dynamics of the source motions does not support a simple scenario of an isotropic radio source and isotropic radio wave scattering \citep[e.g.,][]{1971A&A....10..362S,Arzner1999}; reproducing the observed source dynamics and diagnosing the scattering regime require more complicated scattering simulations. The narrowband ``striae'' bursts are produced, most likely, by small-scale small-amplitude plasma density perturbations in the solar corona connected with propagating MHD waves. These MHD perturbations propagate with the speed larger than the sound speed and closer to Alfv\'en (fast magnetoacoustic) speed in the corona.

\begin{acks}
The work has benefited from a  Marie  Curie  International  Research Staff Exchange Scheme ``Radiosun'' (PEOPLE-2011-IRSES-295272), an international team grant (\url{http://www.issibern.ch/teams/lofar/}) from ISSI Bern, Switzerland, the Program No. 28 of the RAS Presidium, and budgetary funding of Basic Research program II.16. E.P.K. was supported by Science and Technology Facilities Council Grant (STFC) No. ST/P000533/1.
This paper is based (in part) on data obtained from facilities of the International LOFAR Telescope (ILT) under project code LC3-012. LOFAR \citep{Haarlem2013} is the Low Frequency Array designed and constructed by ASTRON. It has observing, data processing, and data storage facilities in several countries, that are owned by various parties (each with their own funding sources), and that are collectively operated by the ILT foundation under a joint scientific policy. The ILT resources have benefitted from the following recent major funding sources: CNRS-INSU, Observatoire de Paris and Universit\'e d'Orl\'eans, France; BMBF, MIWF-NRW, MPG, Germany; Science Foundation Ireland (SFI), Department of Business, Enterprise and Innovation (DBEI), Ireland; NWO, The Netherlands; The Science and Technology Facilities Council, UK.
\end{acks}

\bibliographystyle{spr-mp-sola}

\end{article}
\end{document}